\documentstyle[aaspp4]{article}
\input psfig

\catcode`\@=11 

\def\ergs   {~erg~s$^{-1}$}
\def\ergcms   {~erg~cm$^{-2}$~s$^{-1}$}

\def\sles{\lower2pt\hbox{$\buildrel {\scriptstyle <} 
   \over {\scriptstyle\sim}$}}
\def\sgreat{\lower2pt\hbox{$\buildrel {\scriptstyle >} 
   \over {\scriptstyle\sim}$}}

\begin{document}

\title{Advection-Dominated Accretion and Black Hole Event Horizons}

\authoremail{authors@cfa.harvard.edu}
\author{Ramesh Narayan, Michael R. Garcia, and Jeffrey E. McClintock}

\affil{Harvard-Smithsonian Center for Astrophysics, 60 Garden St.,
Cambridge, MA 02138;
rnarayan@cfa.harvard.edu, mgarcia@cfa.harvard.edu, jmcclintock@cfa.harvard.edu}

\begin{abstract}

The defining characteristic of a black hole is that it possesses an
event horizon through which matter and energy can fall in but from
which nothing escapes.  Soft X-ray transients (SXTs), a class of X-ray
binaries, appear to confirm this fundamental property of black holes.
SXTs that are thought to contain accreting black holes display a large
variation of luminosity between their bright and faint states, while
SXTs with accreting neutron stars have a smaller variation.  This
difference is predicted if the former stars have horizons and the
latter have normal surfaces.

\end{abstract}


\section{Introduction}

Soft X-ray transients (SXTs) are binary stellar systems in which a
black hole (BH) or neutron star (NS) primary accretes matter from a
main sequence or giant secondary (van Paradijs \& McClintock 1995;
Tanaka \& Shibazaki 1996).  A typical SXT displays a large variation
in luminosity.  Most of the time it remains in a quiescent state and
is very dim.  In this phase, only a fraction of the mass transferred
from the secondary accretes on the primary, the rest being stored in
the outer part of an accretion disk.  Once every few decades, however,
the source goes into outburst and becomes very bright for a few
months.  During this time, the material stored in the outer disk is
apparently accreted very rapidly.

In addition to the major outbursts described above, some SXTs also
display Type I bursts due to thermonuclear flashes of the accreted
material, which unambiguously identifies them as NS SXTs (Joss \&
Rappaport 1984).  In a few of these systems the mass of the primary
star has been measured and found to be consistent with the mass of a
NS ($\sim1.4M_\odot$).  In several other SXTs, however, the mass of
the primary is found to be $>3M_\odot$, which makes these stars too
massive to be NSs.  These are identified as BH candidates.

Although BHs and NSs have been convincingly distinguished on the basis
of their masses, we note that the real physical distinction between
the two is that BHs are supposed to have event horizons while NSs are
normal stars with surfaces.  This basic difference between BHs and NSs
has not so far been demonstrated.  We show in this paper that SXTs
provide a unique opportunity to test the reality of event horizons.
Since our argument depends on an understanding of accretion flows, we
begin with a discussion of this subject in \S2.

\section{Models of Accretion Flows}

A standard paradigm in the field of accretion is the thin accretion
disk (Frank, King \& Raine 1992).  In this model, the heat energy
released by viscous dissipation is radiated almost immediately by the
accreting gas, and so the net luminosity is equal to (approximately
one half) the gravitational energy released as the mass falls onto the
central star.  Since the effective radius of a BH, namely the
Schwarzschild radius $R_S$ ($=2GM/c^2$, where $M$ is the mass of the
BH), is not very different from the radius of a neutron star,
$R_{NS}\sim 10~{\rm km}\sim2.5R_S$ (Shapiro \& Teukolsky 1983), the
depth of the gravitational potential is roughly the same for the two
stars.  Therefore, the accretion luminosities are also nearly equal,
corresponding to $\sim10\%$ of the rest mass energy of the accreting
gas, i.e. $L\sim 0.1\dot Mc^2$, where $\dot M$ is the mass accretion
rate.

SXTs in outburst appear to be reasonably well described by the thin
accretion disk model.  The spectra are compatible with this model,
though there are some components in the spectra (e.g. a hard power-law
tail in the case of the BH SXTs) which suggest that a part of 
the accretion may be in a form
other than a thin disk (Tanaka \& Shibazaki 1996).  The maximum
luminosities $L_{max}$ seen in SXTs at the peaks of their outbursts
are $\sim(0.2-1)L_{Edd}$, where $L_{Edd}\sim 10^{38}(M/M_\odot)~{\rm
erg\,s^{-1}}$ is the Eddington luminosity.  This suggests that SXTs
approach the Eddington mass accretion rate $\dot
M_{Edd}\sim1.4\times10^{18}(M/M_\odot)~{\rm g\,s^{-1}}$ in outburst.
BHs in SXTs typically have masses $\sim5-15M_\odot$, which makes them
several times more massive than NSs.  Therefore, we expect larger
values of $L_{max}$ in BH SXTs than in NS SXTs.

BH SXTs in quiescence are not well described by the thin accretion
disk model.  The spectra of two well-observed systems, A0620--00 and
V404 Cyg, are impossible to explain using a thin disk (Narayan,
McClintock \& Yi 1996; Narayan, Barret \& McClintock 1997), and it
appears that a different model is required.

In recent years, considerable work has been done on
advection-dominated accretion flows (ADAFs) which represent a very
different regime of accretion than the thin disk (Narayan \& Yi 1994,
1995; Abramowicz et~al. 1995; Chen et~al. 1995; see Rees et al. 1982
for a discussion of the related ion torus model).  The key feature of
an ADAF is that the radiative efficiency of the accreting gas is low,
so that the bulk of the viscously dissipated energy is stored in the
gas as thermal energy (or entropy).  ADAF solutions exist below a
critical accretion rate, $\dot m<\dot m_{crit} \sim 10^{-2}-10^{-1}$,
where we define $\dot m \equiv \dot M/\dot M_{Edd}$.

The optically thin gas in an ADAF radiates with a spectrum which is
very different from the blackbody-like spectrum of a thin disk, and
the two modes of accretion are easy to distinguish via observations.
More importantly for the purpose of this paper, the luminosity of an
ADAF has a steep dependence on $\dot m$, viz.  $L/L_{Edd}\equiv l \sim
\dot m^2/\dot m_{crit}$ (Narayan \& Yi 1995; Mahadevan 1997).  The
quadratic scaling with $\dot m$ arises because the gas is in the form
of a two-temperature plasma with the ions being much hotter than the
electrons (Shapiro, Lightman \& Eardley 1976).  The efficiency with
which thermal energy is transferred from ions to electrons (to be
subsequently radiated) is proportional to $\dot m$, and therefore
$l\propto\dot m^2$ (Rees et al. 1982).  In contrast, the luminosity of
a thin disk varies as $l \sim\dot m$. The variation of luminosity with
$\dot m$ in the two regimes of accretion is indicated by the solid
line in Fig.~1.  Note that the total energy released by gravity is
always $L_{release}/L_{Edd}\sim\dot m$.  The key difference is that
whereas in a thin disk a large fraction of the released energy is
radiated, in an ADAF nearly all the energy remains locked up in the
gas as thermal energy and is advected into the central star.  If the
star happens to be a BH, the advected thermal energy disappears
through the horizon.

The observed quiescent spectra of the BH SXTs, A0620--00 and V404 Cyg,
are explained well with an ADAF model (Narayan et al. 1996, 1997).
The ADAF extends from the horizon of the BH out to a radius
$\sim10^4R_S$, and a thin disk is present outside this radius.  The
model is consistent with all the spectral information available at
this time.

When an ADAF encounters a BH, an enormous quantity of thermal energy
disappears through the event horizon; the energy flow rate into the
BH is $\sim0.1\dot Mc^2$.  Since the horizon plays such a crucial
role, the success of the ADAF model in the case of A0620--00 and V404
Cyg indicates that these two BH candidates at least are true black
holes with horizons (Narayan et al. 1996, 1997).

What do we expect in the case of a NS SXT?  It is reasonable to assume
that, in quiescence, these systems too undergo accretion via ADAFs.
As in BH SXTs, the accreting gas would radiate very inefficiently, and
the direct luminosity from the accretion flow would be very low
($l\sim\dot m^2/\dot m_{crit}$).  However, when the superheated gas
falls on the NS, the thermal energy does not disappear but rather
heats up the star.  Once a steady state is reached, we would expect
the stellar surface to radiate with a luminosity equal to the rate at
which thermal energy flows into it.  Thus, the total observed
luminosity will be the same as in a thin disk, i.e. $l \sim \dot m$
(Narayan \& Yi 1995), as indicated by the dashed line in Fig. 1.

The comparison shown in Fig. 1 implies that the spread of luminosity
between the outburst and quiescent state in BH SXTs (solid line)
should be significantly larger than the spread in NS SXTs (dashed
line).  In the next section we discuss the observations we have
collected in order to test this prediction.

\section{Observations}

The observations of SXTs listed in Table~1 yield the quiescent
luminosities listed in Table~2.  In order to make an unbiased
comparison of the high and low state luminosities, we have taken the
fluxes from the literature and/or archival data from the HEASARC in
order to compute the emitted luminosities over the 0.5--10.0~keV band.
In computing the quiescent luminosities, we have been motivated by two
points.  First, the only BH~SXT with a well-determined quiescent
spectrum is V404 Cyg (Narayan et~al. 1997).  Its observed spectrum is
a power-law with a photon index $\alpha_N \approx 2.1$ (0.7--8.5~keV),
a result that is in excellent agreement with the prediction of the
ADAF model (Narayan et~al. 1997).  Consequently, in deriving the
quiescent luminosity of A0620--00 and the luminosity limits for the
three fainter BH SXTs, we assume that their spectra also have a
power-law form with $\alpha_N = 2.1$.  Second, the existence of the
neutron star surface gives some physical motivation to a blackbody
spectral shape which we assume (and which is generally consistent with
the data) for the NS SXTs.  For the above spectral models, the bulk of
the luminosity is in the 0.5--10.0 keV band; for example, if the high
energy limit of the band is extended from 10.0 to 100 keV, the
luminosities for both the BH and NS SXTs are increased by $<$ 60\%.

Table~2 shows the maximum luminosities $L_{max}$ and minimum
luminosities $L_{min}$ observed for a number of NS SXTs and BH SXTs.
The luminosities given here require a knowledge of the distances to
the sources, which are somewhat uncertain.  We therefore show in the
final column of Table~2 the quantity $L_{min}/L_{max}$, which is
independent of the distance.  Because the blackbody temperatures of NS
SXTs in quiescence are $\sim 0.3$ keV, the interstellar medium can
absorb a significant fraction of the total flux.  We have corrected
for this absorption for both the NS and BH SXTs using the column
densities (${\rm N_H}$) given in Table 2, which were derived (in most
cases) using the relation of Predehl \& Schmitt (1995) and the
measured values of the optical reddening with A${_V}$ = 3.1E$(B-V)$.
For a few of the NS SXT in outburst a higher ${\rm N_H}$ appears to be
indicated, and was used (see Section~\ref{ind.sources}).  We note that
these outburst luminosities are insensitive to errors in the assumed ${\rm
N_H}$.

\subsection{Notes on Individual Sources}\label{ind.sources}

{\bf BH SXTs:} The values of $L_{max}$ correspond to the energy range
1--40 keV (Tanaka \& Shibazaki 1996) except for H1705--25, for which the
energy range is 2--200 keV (Wilson \& Rothschild 1983).  All of the
luminosity limits are at the $3\sigma$ level of confidence. 
{\bf GS2000+25:} We have combined three observations from the HEASARC
database for a total exposure of 26.5~ksec.  For E$(B-V) = 1.5$ (e.g.,
Chevalier \& Ilovaisky 1990), we find ${\rm N_H} = 8.3 \times
10^{21}$cm$^{-2}$, which is substantially higher than the value
assumed by Verbunt et~al. (1994).

{\bf NS SXTs:} The distances and values of the interstellar column,
${\rm N_H}$, are as summarized in Table~2, unless explicitly mentioned
below.
{\bf X1608--52:} 
Asai et~al. (1996) have measured a quiescent flux with
ASCA which corresponds to a detected luminosity of $5
\times 10^{32}$\ergs (0.5--10~keV) and an 
emitted luminosity of $1.9 \times
10^{33}$\ergs (0.5--10~keV).  Tenma has observed 2--20~keV luminosities
as high as $3 \times 10^{37}$\ergs ~(Mitsuda et~al. 1989).  The
high-state X-ray absorption is measured as $\sim 2 \times
10^{22}$~cm$^{-2}$, and may be higher than that observed in quiescence
due to local absorption.  The emitted 0.5--10~keV luminosity for the
spectral shape and absorption quoted in Mitsuda et~al. (1989) is $1.1
\times 10^{38}$\ergs.
{\bf \hbox{Cen X--4:}} 
The intrinsic, quiescent 0.5--10~keV luminosity has been
measured with ASCA as $2.4 \times 10^{32}$\ergs 
(Asai et~al. 1996).  
During its brightest transient outburst (excluding the peaks of
type~I X-ray bursts), Cen X--4 reached $\sim 1000$ c~s$^{-1}$
in the Ariel~5 ASM, which covered an energy range of 3--12~keV (Evans,
Belian \& Conner 1970).  The corresponding 0.5--10~keV luminosity
assuming a bremsstrahlung spectrum with $kT = 3.9$~keV is $1.2 \times
10^{38}$\ergs.
{\bf Aql X--1:} 
For the observed PSPC count rate of
0.03~c~s$^{-1}$ and a 0.3~keV black body spectrum (Verbunt
et~al. 1994), we calculate a 0.5--10~keV luminosity of $4.4 \times
10^{32}$\ergs.  The brightest outburst of
Aql X--1 recorded to date was observed with Ariel~5, SAS--3, and
Copernicus in 1978 (Charles et~al. 1980). The temperature measured
with Copernicus was $\sim 9$~keV.  We calculate an unabsorbed
0.5--10~keV luminosity of $3.6 \times 10^{37}$\ergs.
{\bf EXO 0748--676:} 
In quiescence, Parmar et~al. (1986) found an intrinsic
0.2--3.5~keV luminosity of $1.2 \times 10^{34}$\ergs\/
and Garcia \& Callanan (1997) determined a blackbody
temperature of 0.2~keV.  
The intrinsic 0.5--10~keV luminosity is $ 1.2 \times
10^{34}$\ergs.  The highest flux yet recorded from EXO~0748--676 is
$1.5 \times 10^{-9}$\ergcms, 1--20~keV (Parmar et~al. 1986).  
Assuming the spectra found by Parmar et~al., we compute a
0.5--10~keV intrinsic luminosity of $3.3 \times 10^{37}$\ergs.
{\bf 4U~2129+47:} 
Using a blackbody temperature of 0.22~keV (Garcia
\& Callanan 1997), we calculate an intrinsic 0.5--10~keV luminosity
of $5.9 \times 10^{32}$\ergs\/ in quiescence.  Calculation
of the intrinsic high state flux is complicated by the facts that (1)
only a small fraction of the flux emitted by the neutron star is
scattered into our line of sight (McClintock et~al. 1982, White \&
Holt 1982), and (2) the mean on-state flux decreased steadily since
discovery (Pietsch et~al. 1986).  The 0.5--10 keV high state
luminosity computed assuming the two-component spectrum in Garcia
(1995) is $9.1 \times 10^{35}$\ergs.  Correcting for a scattering
fraction of 4\% (as measured in EXO0748--676, Parmar et~al. 1986) and
a factor of $\sim 7$~decrease in mean flux since discovery, we find
a maximum intrinsic 0.5--10~keV luminosity of $1.6 \times
10^{38}$\ergs.

\section{Conclusions}

Figure~2 is a plot of $L_{min}/L_{max}$ {\it vs.} $L_{max}$ of the BH
SXTs and NS SXTs discussed in \S3.  We see a clear confirmation of the
basic ideas described in \S2.  First, the BH SXTs all have larger
values of $L_{max}$ than the NS SXTs, consistent with their larger
masses.  Also, none of the NS SXTs has $L_{max}>L_{Edd}=2\times10^{38}
~{\rm erg\,s^{-1}}$ (taking $M_{NS}=1.4M_\odot$).  These facts have
been noted before (e.g. Barret et al. 1996).

Second, and this is new, without exception every BH SXT in Fig.~2 has
a smaller value of $L_{min}/L_{max}$ than every NS SXT.  This is
exactly what we expect if (1) BHs have horizons and NSs do not, and
(2) quiescent SXTs accrete via ADAFs.  Note that for some of the BH
SXTs we only have upper limits on the quiescent luminosity.  When
these fluxes are ultimately measured, the difference between NS SXTs
and BH SXTs may become even more dramatic.

Could the difference in $L_{min}/L_{max}$ merely mean that NS SXTs
experience a smaller range of $\dot m$ between quiescence and outburst
compared to BH SXTs?  This is unlikely since the two systems are very
similar in many respects.  Furthermore, NS SXTs are, if at all, likely
to have a {\it larger} swing of $\dot m$ than BH SXTs.  If a NS has a
strong enough magnetic field and spins rapidly enough, the field can
cause the accreting matter to be flung out through a ``propeller
effect'' (Illarionov \& Sunyaev 1975), thereby dimming the source.
Asai et~al. (1996) and Tanaka \& Shibazaki (1996) have argued for the
propeller effect in the NS SXT Cen X--4 in quiescence.  If many
quiescent NS SXTs undergo the propeller effect, the swing of
luminosity between $L_{min}$ and $L_{max}$ will be enhanced in NS SXTs
as a class.  BHs, on the other hand, cannot have a propeller effect
because of the ``no-hair'' theorem which rules out a permanent
magnetic field on a BH (Shapiro \& Teukolsky 1983).  Since the
presence of the propeller effect in quiescent NS SXTs would tend to
wash out the difference in $L_{min}/L_{max}$ which we predict between
NS SXTs and BH SXTs, the fact that the predicted difference is seen
clearly in the data is especially significant.

Could BH SXTs appear less luminous by expelling most of their energy
through outflows?  This is possible in principle, but is somewhat
contrived since we need a mechanism that is extremely sensitive to
mass, switching on suddenly for accretors more massive than
$1.4M_\odot$.

We suggest that the most natural explanation for the difference in
luminosity swing between BH SXTs and NS SXTs is that BHs have event
horizons and NSs do not.  It is a basic property of a BH event horizon
that it will hide any thermal energy which falls through it.  A NS, on
the other hand, does not have a horizon and must re-radiate whatever
thermal energy it accretes.  We suggest that Figure 2 confirms this
difference.  The argument presented here is robust since it makes use
of one of the most basic observables in astronomy, namely the total
received flux.

\acknowledgments
This work was partially supported by NASA grants NAGW-4269 and NAG
5-2837, contract NAS8-30751, and the Smithsonian Institution Scholarly 
Studies Program.  This research has made use of data obtained through 
the HEASARC Online Service, provided by the NASA/Goddard Space Flight 
Center.

\newpage

\begin{table*}[h]
\begin{center}
\centerline{Table 1}
\vskip 12pt
\centerline{Log of Observations of SXTs in Quiescence}
\vskip 20pt
\begin{tabular}{|lllc|}
\hline
Object & Detector & Date of Observation  &  Time (ksec) \\
\hline 
\multicolumn{4}{|c|}{Neutron Star SXTs}  \\ 
\hline
EXO0748--676 & 	Einstein IPC&	1980 March 22&			5.7      \\
Aql X--1 & 	ROSAT PSPC&	1992 October 15--17&		14.4	          \\
Cen X--4 & 	ASCA GIS+SIS&	1994 February 27--28&		28.0     \\
4U2129+47 &	ROSAT PSPC&	1994 June 3&			30.0	          \\
H1608--522 & 	ASCA GIS+SIS & 	1993 August 12--13&		32.0      \\
\hline
\multicolumn{4}{|c|}{Black Hole SXTs} \\
\hline
H1705--25 &  	ROSAT HRI &     1991 March 19--20      &  	1.6  \\
GS2000+251 & 	ROSAT PSPC &    1993 October 13&        	6.8  \\
GS2000+251 & 	ROSAT PSPC &    1993 April 9--11&        	7.0\\
GS2000+251 & 	ROSAT PSPC &    1992 May 1--7&	        	12.7\\
Nov Mus 91 & 	ROSAT PSPC &    1992 March 1--3 &        	16.7 \\
A0620--00 & 	ROSAT PSPC  &   1992 March 10, 24--27&   	29.8  \\
V404 Cyg & 	ASCA GIS+SIS &  1994 May 9--10 &         	40.0 \\
\hline
\end{tabular}
\end{center}
\end{table*}

\begin{table*}[h]
\begin{center}
\centerline{Table 2}
\vskip 12pt
\centerline{Luminosities of SXTs in Quiescence and Outburst}
\vskip 20pt
\begin{tabular}{|lllrrc|}
\hline
Object & $D({\rm kpc})$ & log($N_H$) & $\log(L_{min})$ &  $\log(L_{max})$ & 
$\log(L_{min}/L_{max})$ \\
\hline
\multicolumn{6}{|c|}{Neutron Star SXTs}  \\
\hline
EXO0748--676 & 	10.0$^1$ 		&22.35$^{2}$
							       &   34.1   &   	37.5   &	 --3.4 \\
Aql X--1 & 	 	~2.5$^3$  	&21.32$^{3,4}$
							       &   32.6   &       37.6   &         --5.0 \\
Cen X--4 & 	 	~1.2$^5$
					&20.85$^{6}$
							       &   32.4   &       38.1   &         --5.7 \\
4U2129+47 &	 	~6.3$^7$  	&21.20$^{7}$
							       &   32.8   &	38.2   &	 --5.4 \\
H1608--522 &		~3.6$^5$
					&22.00$^{5,8}$
							       &   33.3   &       38.0   &         --4.7 \\
\hline
\multicolumn{6}{|c|}{Black Hole SXTs} \\
\hline
H1705--25 &  	8.6$^{9~}$
					   	&21.44$^{10}$
							 	   &  $<$33.7    &     38.3     &       $<$-4.6   \\
GS2000+251 & 	2.7$^{9~}$
					   	&21.92$^{10}$
							       &	$<$32.3    &     38.4     &       $<$-6.1 \\
Nov Mus 91 & 	6.5$^{9~}$
					  	&21.21$^{10}$
							       &	$<$32.6    &     39.1     &       $<$-6.5 \\
A0620--00 &   	1.2$^{9~}$
					   	&21.29$^{10}$
							       &	31.0     &    	38.4    &        --7.4  \\
V404 Cyg &   	3.5$^{11}$
					   	&22.04$^{12}$
							       &	33.2     &     39.3     &        --6.1 \\
\hline 
\end{tabular}
\vskip 0.5in
\begin{minipage}[h]{6.0in}
{\small  
References for Table 2:
[1] Parmar et~al. (1986) 
[2] Schoembs \& Zoeschinger (1990), 
[3] Thorstensen, Charles \& Bowyer (1978), 
[4] Verbunt et~al. (1994), 
[5] Asai et~al. (1996), 
[6] Blair et~al. (1984), 
[7] Cowley \& Schmidtke (1990),
[8] Grindlay \& Liller (1978), 
[9] Barret, McClintock \& Grindlay (1996), 
[10] van Paradijs \& McClintock (1995), 
[11] Tanaka \& Shibazaki (1996),
[12] Narayan et~al. (1997)
}
\end{minipage}
\end{center}
\end{table*}

\clearpage
\newpage

\noindent{\bf Figure Captions}

\bigskip
\noindent
Figure 1. Expected variation of the luminosities of BH SXTs and NS
SXTs as functions of the Eddington-scaled mass accretion rate $\dot m$
(based on Fig. 11 of Narayan \& Yi 1995).  The solid line corresponds
to a $10M_\odot$ BH.  For $\dot m>0.1$, the accretion is assumed to
take place via a thin disk with the standard 10\% efficiency, while
for $\dot m<0.1$, the flow is assumed to be an ADAF with reduced
efficiency.  The dashed line corresponds to a NS SXT
($M_{NS}=1.4M_\odot$), for which the efficiency is taken to be $10\%$
regardless of the mode of accretion.  Since $\dot m$ varies by a few
orders of magnitude between outburst and quiescence, we expect BH SXTs
to exhibit a substantially larger variation of $L$ than NS SXTs.
Figure~2 confirms this prediction.

\noindent 
Figure~2.  The outburst luminosities $L_{max}$ and luminosity ranges
$L_{min}/L_{max}$ of the NS SXTs (open circles) and BH SXTs (filled
circles) listed in Table~2.  Upper limits are indicated by arrows.  We
have left the BH~SXT H1705--25 off the plot because the observation
used to compute the upper limit on $L_{min}$ is more than an order of
magnitude less sensitive than the observations of the other 4~BH~SXTs
(see Table~1).  The vertical dashed line represents $L_{Edd}$ for a
$1.4M_\odot$ NS.  The horizontal dashed line is drawn to emphasize the
point that every BH~SXT has a lower value of $L_{min}/L_{max}$ than
every NS~SXT.  In the context of the ADAF model, this separation of
BH~SXTs and NS~SXTs confirms that BH~SXTs have horizons and NS~SXTs do
not.

\newpage\newpage
\centerline{\psfig{figure=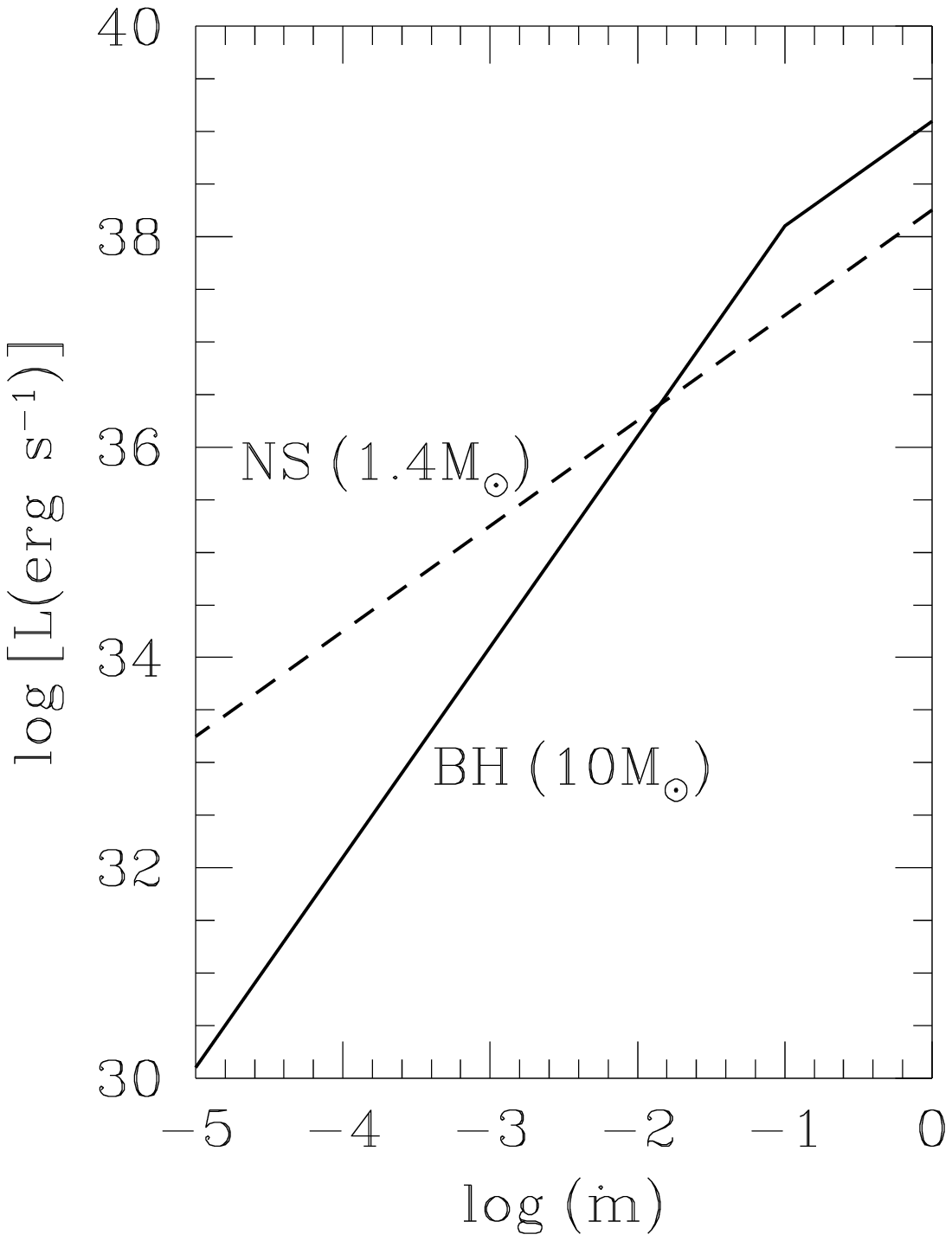,height=13cm,width=12cm}}
\centerline{Figure 1}

\newpage
\centerline{\psfig{figure=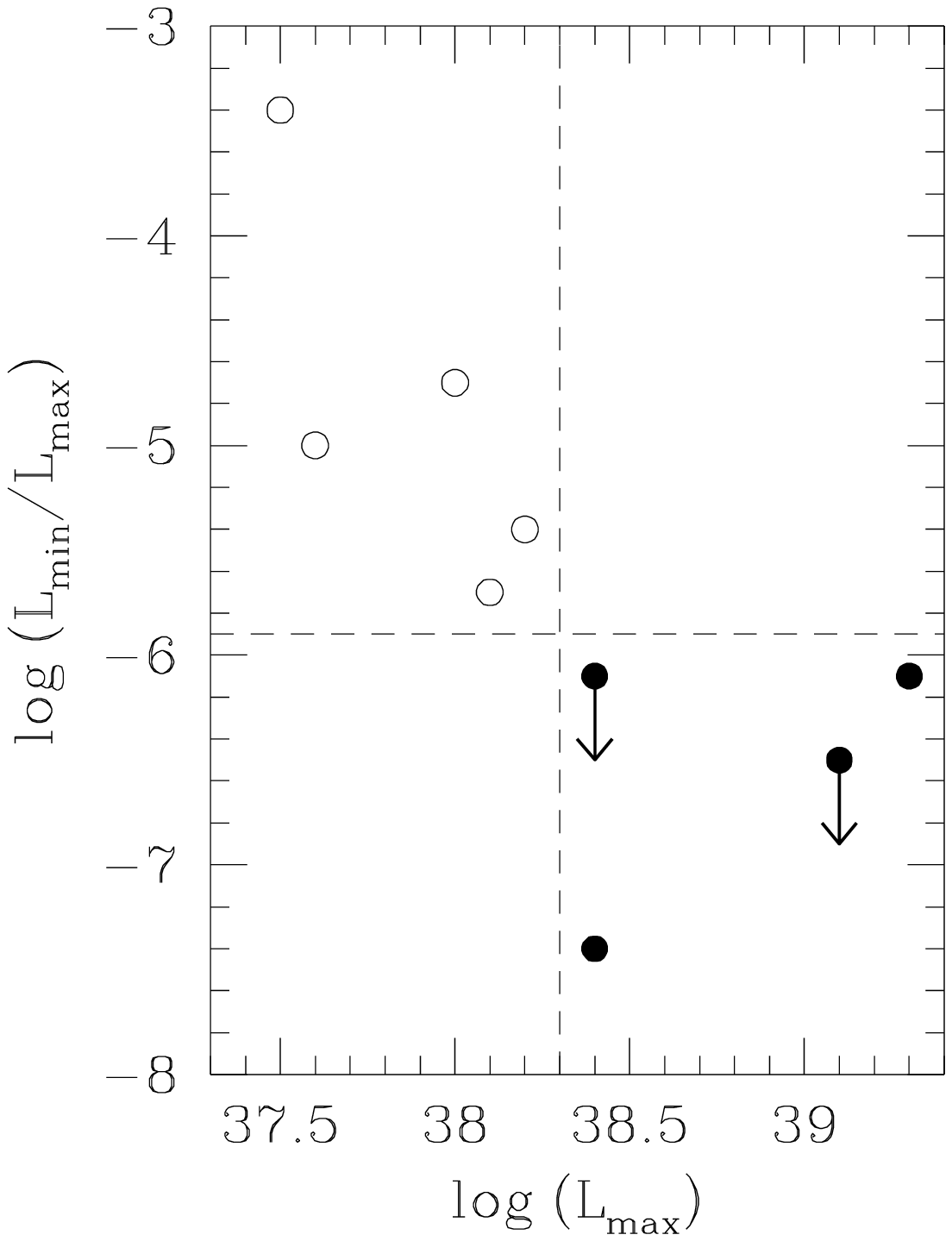,height=13cm,width=12cm}}
\centerline{Figure 2}

\end{document}